\documentclass[]{elsarticle}

\def\R{\mathcal{R}} 
\def\Q{\mathcal{Q}}

\title{\bf Predicting template-based catalysis rates in a simple catalytic reaction model}

\author[unil]{Wim Hordijk\corref{cor}}
\ead{wim@WorldWideWanderings.net}

\author[uc]{Mike Steel}
\ead{mike.steel@canterbury.ac.nz}

\cortext[cor]{Corresponding author. Email: {\tt wim@WorldWideWanderings.net}, phone: +41216924268.}
\address[unil]{University of Lausanne, Dept. of Ecology and Evolution, 1015 Lausanne, Switzerland}
\address[uc]{University of Canterbury, Biomathematics Research Centre, Private Bag 4800, Christchurch, New Zealand}

\begin{document}

\begin{abstract}
We show that in a particular model of catalytic reaction systems, known as the binary polymer model, there is a mathematical invariance between two versions of the model: (1) random catalysis and (2) template-based catalysis. In particular, we derive an analytical calculation that allows us to accurately predict the (observed) required level of catalysis in one version of the model from that in the other version, for a given probability of having self-sustaining autocatalytic sets exist in instances of both model versions. This provides a tractable connection between two models that have been investigated in theoretical origin-of-life studies.
\end{abstract}


\maketitle

\bigskip
\section{Introduction}

In previous work, we introduced and analyzed a model of  catalytic reaction systems (CRS) \cite{Steel:00,Hordijk:04,Mossel:05,Hordijk:11}, based on the original model of Kauffman \cite{Kauffman:86,Kauffman:93}. This model consists of molecule types (bit strings up to a given length $n$), reactions (ligation and cleavage), and randomly assigned catalysis, where molecule types can catalyze reactions with a certain (given) probability. Such models were developed and investigated within the context of theoretical origin-of-life studies. In this setting, a question of particular interest is the level of catalysis required to have a high probability of {\it autocatalytic sets} (``closed'', self-sustaining subsets of molecules and reactions) appearing in instances of this random binary polymer CRS model. We showed, both theoretically and computationally, that this level (in terms of the average number of reactions catalyzed by any one molecule) needs only grow linearly with $n$ (the maximum molecule size in the system) to get a high probability of autocatalytic sets \cite{Hordijk:04,Mossel:05}. However, in \cite{Hordijk:11}, we showed that there is still some discrepancy between the theoretically predicted and computationally observed required levels of catalysis, and we provided a partial explanation for this discrepancy. The theoretical analysis thus provides an important (linear) upper bound, but not (yet) an exact prediction.

In \cite{Hordijk:11}, we also analyzed an extension of the basic model, based on initial experiments reported in \cite{Kauffman:93}, where instead of completely randomly assigned catalysis, a molecule has to match the reaction template (a certain number of bits around the reaction site) to be able to be a catalyst for that particular reaction. For this more realistic version of the CRS model, we also showed (again both theoretically and computationally) that a linear growth rate (in $n$) in the level of catalysis suffices to get a high probability of autocatalytic sets. However, in this case there is also a discrepancy between the theoretically predicted and computationally observed values.

In this paper we take a different approach and ask whether it is possible to accurately predict the (observed) required level of catalysis in the template-based version of the model given the (observed) values in the original, completely random version of the model (and vice versa). At first, this may not seem obvious, as the two versions of the model have quite different constraints on which molecules can/will catalyze which reactions. However, we derive an analytical calculation that allows us to make  this prediction  to a very high degree of accuracy. Therefore, we can conclude that, as far as the probability of the existence of self-sustaining autocatalytic sets is concerned, the more realistic template-based version of the model is (mathematically) no different from the simpler original (completely random) version of the model.

The outline of this paper is as follows. In the next section, we briefly review the mathematical CRS model, in particular the relevant notation and definitions, and highlight the differences between the two versions of the model. In Section \ref{sec:match-prob} we derive, step by step, an analytical calculation that allows us to directly relate the required level of catalysis in both model versions. In Section \ref{sec:predict}, we then show the close match between the analytically predicted and actually observed required level of catalysis in the template-based model. Finally, Section \ref{sec:conclusion} summarizes the main results and conclusions.

\section{A model of catalytic reaction systems} \label{sec:model}

Consider a {\it catalytic reaction system} (CRS) $\Q = (X,\R,C)$, where $X$ is a set of molecule types $x$, $\R$ is a set of reactions $r$ (converting reactants into products) and $C$ is a catalysis set, i.e., a set of molecule-reaction pairs $(x,r)$ indicating that molecule $x$ can catalyze reaction $r$. We also include the notion of a {\it food set} $F \subset X$, which is assumed to contain molecule types that are freely available in the environment. See \cite{Hordijk:04,Steel:00} for a detailed definition.

A particular CRS model that was introduced in \cite{Kauffman:86,Kauffman:93}, here referred to as the {\it binary polymer model}, consists of:
\begin{itemize}
\item A set of molecule types represented by bit strings up to a given length $n$, i.e.,  $X = \{0,1\}^{\leq n}$;
\item A food set consisting of bit strings up to a given (small) length $t < n$, i.e.,  $F = \{0,1\}^{\leq t}$;
\item Two types of reactions: (1) ligation, which ``glues'' two smaller bit strings together into one larger bit string, and (2) cleavage, which breaks a larger bit string into two smaller ones. The reaction set $\R$ consists of all such reactions that are possible within the constraint of the maximum molecule length $n$.
\item Randomly assigned catalysis, where each possible molecule-reaction pair $(x,r)$ is independently assigned to the set $C$ with a given probability $p(n)$ (this probabilistic assignment is done once at the model instantiation; each such instantiation thus gives rise to a different set $C$).
\end{itemize}
Thus,  $n$, $t$, and $p(n)$ are the model parameters.

\begin{figure}[htb]
\centering
\includegraphics[width=12cm]{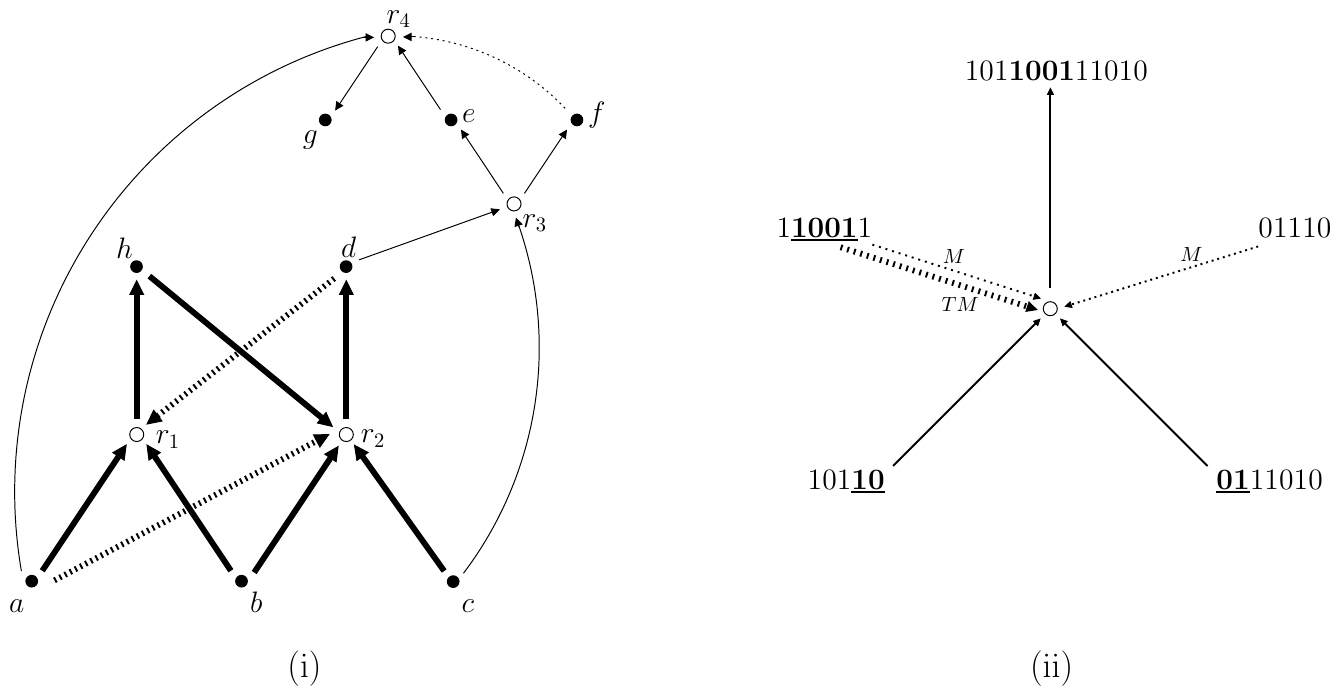}
\caption{ (i) An example of a simple CRS in which $a,b$ and $c$ constitute the food set, catalysis is indicated by dashed lines, and where the reactions $\R'=\{r_1, r_2\}$ (shown in bold) form an RAF set (each reaction in $\R'$ is catalyzed by at least one molecule that is either in the food set or can be produced from it by repeated application of reactions only from $\R'$, and all reactants of the reactions in $\R'$ are either in the food set or can be produced from it by reactions from $\R'$ only). This is the only RAF present in this CRS.
(ii) In the original polymer model ($M$), either one of the potential catalysts shown (110011 and 01110) has the same probability $p$ of catalyzing the ligation reaction $10110 + 0111010  \rightarrow 101100111010$; in the template-based model ($TM$), only 110011 contains a matching template (namely *1001*) for this reaction. In order for the two models to have equal probabilities of containing an RAF, the probability $p'$ that a template-matching molecule catalyzes a reaction needs to be elevated in the $TM$ model (indicated by the bold dashed line) by a factor $m$ (relative to $p$ in model $M$) that can be calculated analytically.}
\label{fig:graph0}
\end{figure}

The main question that was studied in \cite{Kauffman:86,Kauffman:93} with this model is: under which conditions (model parameter values) is there a high probability of an autocatalytic set existing within a full CRS? An {\it autocatalytic set} can be (informally) described as a subset of reactions $\R' \subset \R$ (and the molecules involved in these reactions) in which (1) each reaction $r \in \R'$ is catalyzed by at least one molecule that is either in the food set or can be produced from it by repeated application of reactions only from $\R'$, and (2) all reactants of the reactions $r \in \R'$ are either in the food set or can be produced from it by reactions from $\R'$ only (a simple example is provided in Fig.~\ref{fig:graph0}(i)). For a formal definition, see \cite{Hordijk:04}, where we used the term RAF (reflexively autocatalytic and food-generated) for such autocatalytic (sub)sets.

Note that this CRS and RAF formalism is similar, or at least related, to alternative models in the context of the origin of life, such as petri nets \cite{Sharov:91}, (M,R) systems \cite{Rosen:91,Letelier:06,Jaramillo:10}, the chemoton model \cite{Ganti:97}, other artificial chemistries and topological approaches \cite{Benkoe:09}, and several other frameworks (see also \cite{Hordijk:10,Letelier:11} for a more complete overview and comparison). However, what most of these other formalisms seem to be missing, is some way of actually finding or identifying autocatalytic sets within a larger catalytic reaction system, and a thorough analysis of the probabilities of finding such subsets under different conditions (model parameters).

In \cite{Hordijk:04}, we introduced an efficient algorithm to find RAF sets in any (general) CRS, and applied it to instances of the binary polymer model. We showed both computationally \cite{Hordijk:04} and theoretically \cite{Mossel:05} that a linear growth rate (with $n$, the maximum molecule size) in the level of catalysis (in terms of the average number of reactions catalyzed per molecule) is sufficient to achieve a high probability $P_n$ of autocatalytic sets occurring in instances of the random binary polymer CRS model.

In \cite{Hordijk:11}, we then analyzed a chemically more realistic version of the binary polymer model: template-based catalysis. This differs from the original model in the way molecule-reaction pairs $(x,r)$ are included in the catalysis set $C$:
\begin{itemize}
\item Each possible molecule-reaction pair $(x,r)$ is {\it considered} for inclusion in the set $C$ with a given probability $p'(n)$, but is only {\it allowed} to be included if the molecule $x$, somewhere along its length, matches the {\it reaction template} of reaction $r$.
\end{itemize}
The reaction template, following \cite{Kauffman:93}, is made up of the two bits on either side of the reaction site, or four bits in total (in previous studies, the {\it complement} of the reaction template was actually used; however, given that the model uses bit strings, this does not make a difference in terms of the mathematics due to the inherent symmetry). So, for example, in the ligation reaction $000 + 11111 \rightarrow 00011111$, the reaction template is $0011$, i.e.,   the last two bits of the first reactant plus the first two bits of the second reactant (another example is provided in Fig.~\ref{fig:graph0}(ii)). For this template-based catalysis version of the model, we also showed that a linear growth rate (with $n$) in the level of catalysis suffices for autocatalytic sets to appear with high probability (although with a higher coefficient, or slope, in the linear relation than for the original model).

So, in the original (template-free) random catalysis version of the model, we have $\Pr[(x,r) \in C] = p(n)$. Let $m(n)$ be the probability that an arbitrary molecule (of length at most $n$) matches the four-bit reaction template of an arbitrary reaction.  Then  $\Pr[(x,r) \in C] = p'(n) \cdot m(n)$ in the template-based version of the model.  This suggest that taking $p'(n) = p(n)/m(n)$ might lead to a similar probability $P_n$ of finding autocatalytic sets in the template-based model as with $p(n)$ in the original model. We will show that this heuristic estimate turns out to be extremely accurate, which is not clear {\em a priori}, as the two models are quite different (in the original model molecules catalyze each reactions with identical probabilities, while in the template-based model, these probabilities are highly heterogeneous). So, to predict the required level of catalysis $p'(n)$ in the template-based version of the model, given $p(n)$ and a particular value of $P_n$, we need to know the probability $m(n)$.

\section{The molecule-reaction template match probability} \label{sec:match-prob}

In this section, we derive a precise mathematical procedure for calculating the molecule-reaction template match probability $m(n)$ analytically. This procedure consists of several steps, which will be explained in detail.  We then show the results of applying the derived procedure to calculate the actual template match probabilities.

\subsection{The number of substring matches}

As the first step, we obtain a procedure to calculate analytically the number $f_s(n)$ of bit strings of a given length $n$ that contain a given substring $s$ of given length. We will describe this in detail for a substring $s$ of length four, but the procedure is applicable for any substring length. In the context of our CRS model, $f_s(n)$ is equivalent to the number of molecules of a given length $n$ that match a given four-bit reaction template $s$.

We apply a mathematical technique, known as the {\it transfer-matrix method} \cite{Stanley:86}, which calculates the number $\overline{f_s}(n)$ of bit strings of length $n$ that {\it do not} contain a given substring $s$. This is actually a very general method with a wide range of applicability, but here it will be explained explicitly in the context of counting bit strings of length $n$ that contain a given substring of length four.

\begin{figure}[htb]
\centering
\includegraphics[width=4cm]{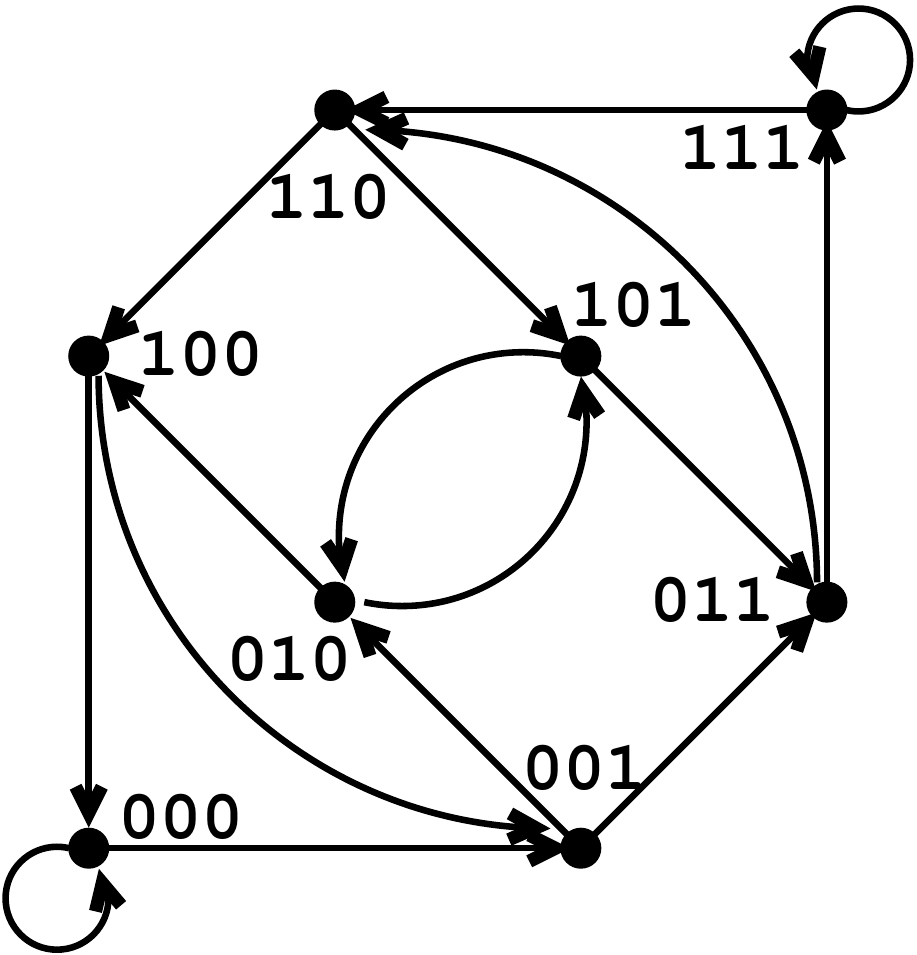}
\caption{The state transition graph $G$ that generates {\it all} possible bit strings.}
\label{fig:graph1}
\end{figure}

To apply this method, we first need to construct a {\it state transition graph} $G_s$ that can generate all bit strings (of any length) that {\it do not} contain a given substring $s$ of length four. For this purpose, it is instructive to  start with a state transition graph $G$ that generates {\it all} possible bit strings. This graph $G$ is shown in Fig.  \ref{fig:graph1}, and can be interpreted as follows.
Each node represents one of eight possible states, with a state being defined (and labeled) by the last three bits that were generated at any given time step. For example, the bottom-left node labeled $000$ represents a state where the last three generated bits were all zero. The bit that will be generated at the next time step can either be another zero (a state transition represented by the arrow that loops back to the same node), or a one (a state transition represented by the arrow that points to the node labeled $001$). Following {\it allowed} state transitions (arrows) this way, all possible bit strings can be generated. This state transition graph $G$ can be represented mathematically by its {\it adjacency matrix} $A$:
\begin{equation}
 A = \bordermatrix{ ~  & 000 & 001 & 010 & 011 & 100 & 101 & 110 & 111 \cr
   000 &  1  &  1  &  0  &  0  &  0  &  0  &  0  &  0  \cr
   001 &  0  &  0  &  1  &  1  &  0  &  0  &  0  &  0  \cr
   010 &  0  &  0  &  0  &  0  &  1  &  1  &  0  &  0  \cr
   011 &  0  &  0  &  0  &  0  &  0  &  0  &  1  &  1  \cr
   100 &  1  &  1  &  0  &  0  &  0  &  0  &  0  &  0  \cr
   101 &  0  &  0  &  1  &  1  &  0  &  0  &  0  &  0  \cr
   110 &  0  &  0  &  0  &  0  &  1  &  1  &  0  &  0  \cr
   111 &  0  &  0  &  0  &  0  &  0  &  0  &  1  &  1  \cr}
\label{eq:adj1}
\end{equation}
where there is a 1 in position $(i,j)$ if there is an arrow in the corresponding graph $G$ from state $i$ to state $j$;  otherwise, there is a 0.

\begin{figure}[htb]
\centering
\includegraphics[width=4cm]{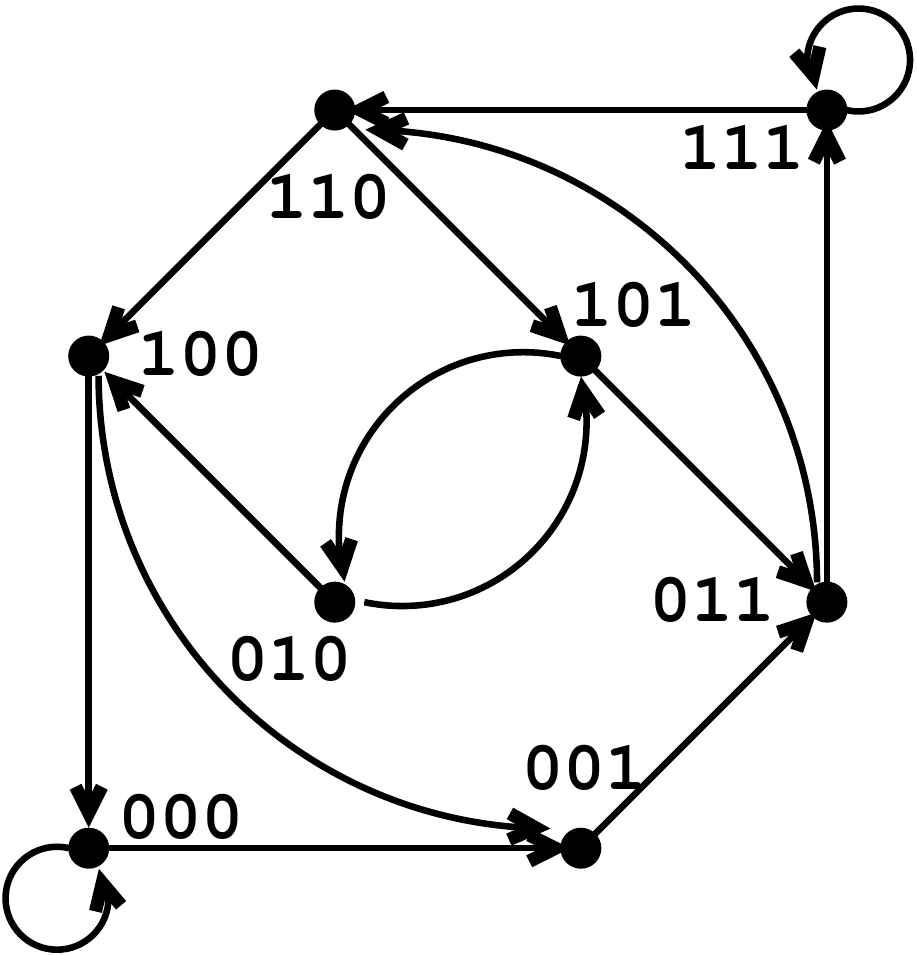}
\caption{The state transition graph $G_{0010}$ that generates all bit strings that {\it do not} contain the substring {\tt 0010}.}
\label{fig:graph2}
\end{figure}

From this general graph $G$ (and matrix $A$), it is now straightforward to construct the corresponding graph $G_s$ (and matrix $A_s$) that generates all bit strings that {\it do not} contain a given substring $s$ of length four. For example, consider $s=0010$. In this case, in the graph $G$ of Fig.  \ref{fig:graph1}, making the transition from node $001$ to node $010$ would not be allowed.  Equivalently, the entry $(001,010)$ in the adjacency matrix $A$ should be set to zero. The resulting graph $G_{0010}$ and adjacency matrix $A_{0010}$ are shown in Fig. \ref{fig:graph2} and Eqn. (\ref{eq:adj2}), respectively (the changed entry in the adjacency matrix is shown in bold).
\begin{equation}
  A_{0010} = \bordermatrix{ ~  & 000 & 001 & 010 & 011 & 100 & 101 & 110 & 111 \cr
   000 &  1  &  1  &  0  &  0  &  0  &  0  &  0  &  0  \cr
   001 &  0  &  0  &  {\bf 0}  &  1  &  0  &  0  &  0  &  0  \cr
   010 &  0  &  0  &  0  &  0  &  1  &  1  &  0  &  0  \cr
   011 &  0  &  0  &  0  &  0  &  0  &  0  &  1  &  1  \cr
   100 &  1  &  1  &  0  &  0  &  0  &  0  &  0  &  0  \cr
   101 &  0  &  0  &  1  &  1  &  0  &  0  &  0  &  0  \cr
   110 &  0  &  0  &  0  &  0  &  1  &  1  &  0  &  0  \cr
   111 &  0  &  0  &  0  &  0  &  0  &  0  &  1  &  1  \cr}
\label{eq:adj2}
\end{equation}
Similarly, for each of the other 15 possible templates $s$ of length four, there is a state transition graph $G_s$ where one of the 16 arrows of the original graph $G$ is left out, and a corresponding adjacency matrix $A_s$ where one of the 16 ones in matrix $A$ is set to zero.

Now, according to the transfer-matrix method \cite{Stanley:86}, the number $\overline{f_s}(n)$ of bit strings of length $n$ that {\it do not} contain a given substring $s$ of length four can be obtained by first computing the following generating function:
\begin{equation}
  F_s(\lambda) = \sum_{i,j=1}^8 \left ( I - \lambda A_s \right ) _{ij}^{-1}
\label{eq:F(l)}
\end{equation}
and then taking the coefficient of $\lambda^{n-3}$ in $F_s(\lambda)$ as the value for $\overline{f_s}(n)$.  Note that, in Eqn. (\ref{eq:F(l)}), $I$ refers to the $8 \times 8$ 
identity matrix, and $\left( I - \lambda A_s \right )_{ij}^{-1}$ is the row $i$ and column $j$ entry of the inverse of the matrix $I- \lambda A_s$.

Given that there are $2^n$ bit strings of length $n$, the number $f_s(n)$ of bit strings of length $n$ that contain a given substring $s$ of length four is now simply:
\begin{equation}
  f_s(n) = 2^n - \overline{f_s}(n).
\label{eq:f(n)}
\end{equation}

\subsection{The template match probability}

As the next step, using the results of Eqn. (\ref{eq:f(n)}), the average (or expected) number $g(n)$ of bit strings of length $n$ that contain an arbitrary substring of length four, is now easily calculated as 
\begin{equation}
  g(n) = \frac{1}{16} \sum_{s \in \{0,1\}^4} f_s(n).
\label{eq:g(n)}
\end{equation}
From this, the probability $h(n)$ that a bit string of length {\it at most} $n$ contains an arbitrary substring of length four, is then calculated as
\begin{equation}
  h(n) = \frac{\sum_{i=1}^n g(i)}{2^{n+1}-2},
\label{eq:h(n)}
\end{equation}
where $2^{n+1}-2$ is the total number of bit strings of length at most $n$. In practice, the summation needs only run from $i=4$ to $n$, as any bit string of length $n < 4$ will obviously not contain a substring of length four (i.e., $g(n) = 0$ for $n < 4$).

In the context of our CRS model, this probability $h(n)$ can be taken as the probability that an arbitrary molecule up to length $n$ matches an arbitrary four-bit reaction template.

\subsection{The fraction of reactions with a four-bit reaction template}

So far, we have assumed that there actually is a substring of length four to be contained (or not) in an arbitrary bit string (or, in our context, a four-bit reaction template to be matched by a molecule). However, in the template-based catalysis version of our model, this is not always the case. In particular, all reactions that have at least one molecule of length one as a reactant (or product, in case of a ligation), do not have a valid four-bit reaction template, given that we require two bits on each side of the reaction site to make up a four-bit template. However, these reactions can still be considered (with probability $p'(n)$), along  with a molecule $x$, for inclusion in the catalysis set $C$ (which, of course, will never actually be ``approved'', given that they do not have a valid four-bit template).

So, as the final step, we need to calculate the fraction $k(n)$ of reactions that actually do have a four-bit reaction template, in order to ``discount'' for those reactions that might be considered, but which do not have a valid reaction template. A straightforward counting argument will provide us with a simple expression for this fraction $k(n)$.

First, recall that the total number of reactions, for a given $n$, is $(n-2)2^{n+1}+4$. Next, we need to count the number of reactions that have at least one molecule of length one as a reactant (since reactions are considered bi-directional, we do not need to consider cleavage reactions separately). A molecule of length one can react (ligate) with any other molecule of length at most $n-1$ (otherwise, the maximum molecule length $n$ would be violated). There are two molecules of length one, which can be either the first or the second reactant (i.e.,  four combinations) and there are $2^n-2$ molecules of length at most $n-1$, so we have  $4(2^n-2)-4$ reactions with at least one molecule of length one as a reactant (the `$-4$' at the end is because otherwise we would double-count the four possible reactions where {\it both} reactants are of length one). So, the fraction $\overline{k}(n)$ of reactions {\it without} a valid four-bit reaction template is:
\begin{equation}
  \overline{k}(n) = \frac{4(2^n-2)-4}{(n-2)2^{n+1}+4} =
  \frac{2^{n+2}-12}{(n-2)2^{n+1}+4} \approx \frac{2^{n+2}}{(n-2)2^{n+1}} =
  \frac{2}{n-2}.
\end{equation}
Consequently, the fraction $k(n)$ of reactions {\it with} a valid four-bit reaction template is
\begin{equation}
  k(n) = 1 - \overline{k}(n) = 1 - \frac{2}{n-2} = \frac{n-4}{n-2}.
\label{eq:k(n)}
\end{equation}
Note that this is an approximation, and the exact number is actually equal to:
\begin{equation}
  k(n) = \frac{n-4 + \frac{1}{2^{n-3}}}{n-2 + \frac{1}{2^{n-1}}}.
\end{equation}
However, even for relatively small values of $n$ (10 or higher suffices for our purposes), the difference between the exact and simpler approximate values becomes negligible.

Now we can finally put everything together and calculate the probability $m(n)$ that an arbitrarily chosen molecule of length at most $n$ will match the four-bit reaction template of an arbitrarily chosen reaction (including the possibly that a reaction does not have a valid reaction template, in which case there is no match), by combining Eqns. (\ref{eq:h(n)}) and (\ref{eq:k(n)}):
\begin{equation}
  m(n) = k(n) \cdot h(n) = \frac{n-4}{n-2}h(n).
\label{eq:m(n)}
\end{equation}

\subsection{Results}

We used the software package {\tt wxMaxima} \cite{wxMaxima}, which is capable of performing symbolic computation and is available for free under the GPL license, to compute the generating functions $F_s(\lambda)$ (Eqn. (\ref{eq:F(l)})) and the corresponding values of $f_s(n)$ (Eqn. (\ref{eq:f(n)})). For example, for $s=0010$ (using the adjacency matrix $A_{0010}$ in Eqn. (\ref{eq:adj2})), we get:
\begin{equation}
  F_{0010}(\lambda) = \frac{4\lambda^3-2\lambda^2-\lambda+8}{-\lambda^4+\lambda^3-2\lambda+1}.
\end{equation}
After applying a Taylor expansion, this yields:
\begin{equation}
  F_{0010}(\lambda) = 8+15\lambda+28\lambda^2+52\lambda^3+97\lambda^4+181\lambda^5+\ldots
\end{equation}
So, for example, $\overline{f_{0010}}(5) = 28$, i.e., the coefficient of $\lambda^{5-3} = \lambda^2$ in $F_{0010}(\lambda)$. From this, it can be easily calculated that $f_{0010}(5) = 2^5 - \overline{f_{0010}}(5) = 32-28 = 4$, i.e., four bit strings of length $n=5$ contain the substring $s=0010$. The results for all templates and $4 \leq n \leq 20$ are given in Table \ref{tab:f(n)}. Note that because of the inherent symmetry, we only show values for the eight templates that start with a $0$. To find the value for a template $s$ that starts with a $1$, simply look up the value for its (binary) complement $\overline{s}$. For example, $f_{1010}(12) = f_{0101}(12) = 1731$.  Notice also that other symmetries exists -- in particular, $f_n(s)$ is identical to $f_n(s')$ when $s'$ is template $s$ in reverse order. This and further symmetries ensure that the number of {\it distinct} columns in Table \ref{tab:f(n)} is only four, rather than the eight possible.

\begin{table}[htb]
\begin{center}
{\small
\begin{tabular}{lrrrrrrrr}
\hline
$n$/$s$ & 0000 & {\bf 0001} & \underline{0010} & {\bf 0011} & \underline{0100} & 0101 & \underline{0110} & {\bf 0111} \\
\hline
4 & 1 & 1 & 1 & 1 & 1 & 1 & 1 & 1 \\
5 & 3 & 4 & 4 & 4 & 4 & 4 & 4 & 4 \\
6 & 8 & 12 & 12 & 12 & 12 & 11 & 12 & 12 \\
7 & 20 & 32 & 31 & 32 & 31 & 28 & 31 & 32 \\
8 & 48 & 79 & 75 & 79 & 75 & 68 & 75 & 79 \\
9 & 111 & 186 & 174 & 186 & 174 & 158 & 174 & 186 \\
10 & 251 & 424 & 393 & 424 & 393 & 357 & 393 & 424 \\
11 & 558 & 944 & 870 & 944 & 870 & 792 & 870 & 944 \\
12 & 1224 & 2065 & 1897 & 2065 & 1897 & 1731 & 1897 & 2065 \\
13 & 2656 & 4456 & 4087 & 4456 & 4087 & 3738 & 4087 & 4456 \\
14 & 5713 & 9512 & 8721 & 9512 & 8721 & 7996 & 8721 & 9512 \\
15 & 12199 & 20128 & 18463 & 20128 & 18463 & 16972 & 18463 & 20128 \\
16 & 25888 & 42287 & 38832 & 42287 & 38832 & 35789 & 38832 & 42287 \\
17 & 54648 & 88310 & 81222 & 88310 & 81222 & 75052 & 81222 & 88310 \\
18 & 114832 & 183492 & 169086 & 183492 & 169086 & 156647 & 169086 & 183492 \\
19 & 240335 & 379624 & 350571 & 379624 & 350571 & 325616 & 350571 & 379624 \\
20 & 501239 & 782497 & 724288 & 782497 & 724288 & 674436 & 724288 & 782497 \\
\hline
\end{tabular}
}
\end{center}
\caption{Values for $f_s(n)$ for all eight possible four-bit templates that start with a $0$, and for $4 \leq n \leq 20$, as calculated using the transfer-matrix method. By symmetry, we only need to describe templates that start with $0$. Other symmetries are also apparent in the table: notice that the three columns headed by a bold template are identical, and the three columns headed by an underlined template are also identical.}
\label{tab:f(n)}
\end{table}

Using the values from Table \ref{tab:f(n)}, we can compute $g(n)$ (using Eqn. (\ref{eq:g(n)})), then $h(n)$ (using Eqn. (\ref{eq:h(n)})) and $k(n)$ (using Eqn. (\ref{eq:k(n)})), and, finally, $m(n)$ (using Eqn. (\ref{eq:m(n)})). The results of these calculations are shown in Table \ref{tab:all} for $4 \leq n \leq 20$. The (analytically) calculated values for $f_s(n)$ and $g(n)$ were verified by a small computer program we wrote to explicitly enumerate all possible bit string/template combinations and counting the number of matches. However, for larger values of $n$, this would obviously be intractable, whereas the analytical (transfer-matrix) method will still be feasible.

\begin{table}[htb]
\begin{center}
\begin{tabular}{rrrrr}
\hline
$n$ &    $g(n)$ &    $h(n)$ & $k(n)^{*}$ &   $m(n)$ \\
\hline
 4 &      1.000 &  0.033333 & 0.235294 & 0.007843 \\
 5 &      3.875 &  0.078629 & 0.408163 & 0.032093 \\
 6 &     11.375 &  0.128968 & 0.527132 & 0.067983 \\
 7 &     29.625 &  0.180610 & 0.610592 & 0.110279 \\
 8 &     72.250 &  0.231618 & 0.671001 & 0.155416 \\
 9 &    168.625 &  0.280577 & 0.714695 & 0.200527 \\
10 &    382.375 &  0.327041 & 0.750000 & 0.245281 \\
11 &    849.000 &  0.370817 & 0.777778 & 0.288413 \\
12 &   1855.125 &  0.411874 & 0.800000 & 0.329499 \\
13 &   4002.875 &  0.450258 & 0.818182 & 0.368393 \\
14 &   8551.000 &  0.486087 & 0.833333 & 0.405072 \\
15 &  18118.000 &  0.519503 & 0.846153 & 0.439579 \\
16 &  38129.250 &  0.550655 & 0.857142 & 0.471990 \\
17 &  79787.000 &  0.579691 & 0.866667 & 0.502399 \\
18 & 166151.625 &  0.606755 & 0.875000 & 0.530911 \\
19 & 344567.000 &  0.631982 & 0.882353 & 0.557631 \\
20 & 712003.750 &  0.655501 & 0.888889 & 0.582668 \\
\hline
\end{tabular}
\end{center}
\caption{The analytically calculated values for $g(n)$, $h(n)$, $k(n)$, and $m(n)$, for $4 \leq n \leq 20$. $^{*}$For the calculation of $k(n)$, we used the exact formula for $n < 10$ and the approximation for $n \geq 10$. For $n=10$, the two values differ by only 0.1\%.}
\label{tab:all}
\end{table}

\section{Predicting the required level of catalysis} \label{sec:predict}

Table \ref{tab:all} provides us with the molecule-reaction template match probabilities $m(n)$ which can be used to predict the required level of $p'(n)$ (the probability of considering a molecule-reaction pair for inclusion in the catalysis set $C$), given $p(n)$ and a probability $P_n$ of observing autocatalytic sets.

To test these predictions, we performed computer simulations with the two versions of the binary polymer model, using our RAF algorithm to detect autocatalytic sets in instances of these CRS models. For various values of $n$, we identified the required levels of catalysis $p(n)$ (random catalysis model) and $p'(n)$ (template-based model) for which there was a probability $P_n=0.5$ of finding RAF sets\footnote{In previous papers, we have already shown that $P_n=0.5$ provides a useful reference point.}. In the template-based catalysis version of the model, we used reaction templates of four bits (two bits on each side of the reaction site) that have to be matched completely by a candidate catalyst (no partial matches allowed). We used a value of $t=4$ for the food set (i.e., molecules up to length four) to allow at least some food molecules to also be catalysts. For each combination of $n$ and $p(n)$ (or $p'(n)$), we generated between 500 (for larger $n$) and 10,000 (for smaller $n$) instances of the binary polymer model, and applied our RAF algorithm to count the fraction ($P_n$) of these instances that contain an RAF set. Due to the exponential growth in the size of the reaction set $\R$ with increasing $n$, we went up to $n=18$ only, which already took many weeks of computing time, even on a large parallel computer cluster.

\begin{figure}[htb]
\centering
\includegraphics[width=8cm, angle=-90]{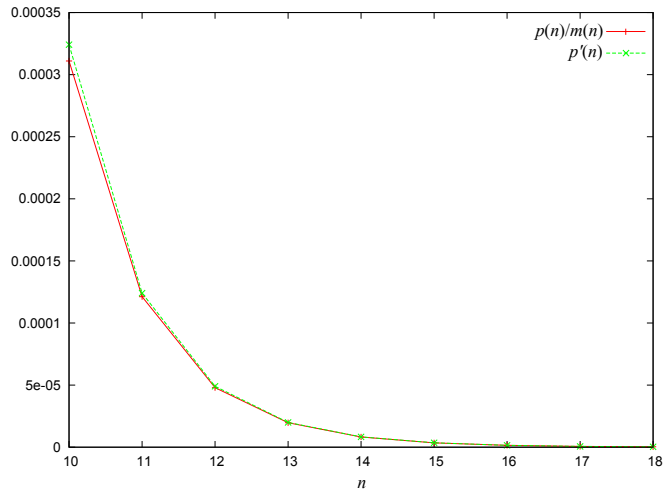}
\caption{Comparison of the predicted ($p(n)/m(n)$; solid line) and observed ($p'(n)$; dotted line) values of the required level of catalysis in the template-based version of the binary polymer CRS model, for various values of $n$.}
\label{fig:compare}
\end{figure}

\begin{figure}[htb]
\centering
\includegraphics[width=8cm, angle=-90]{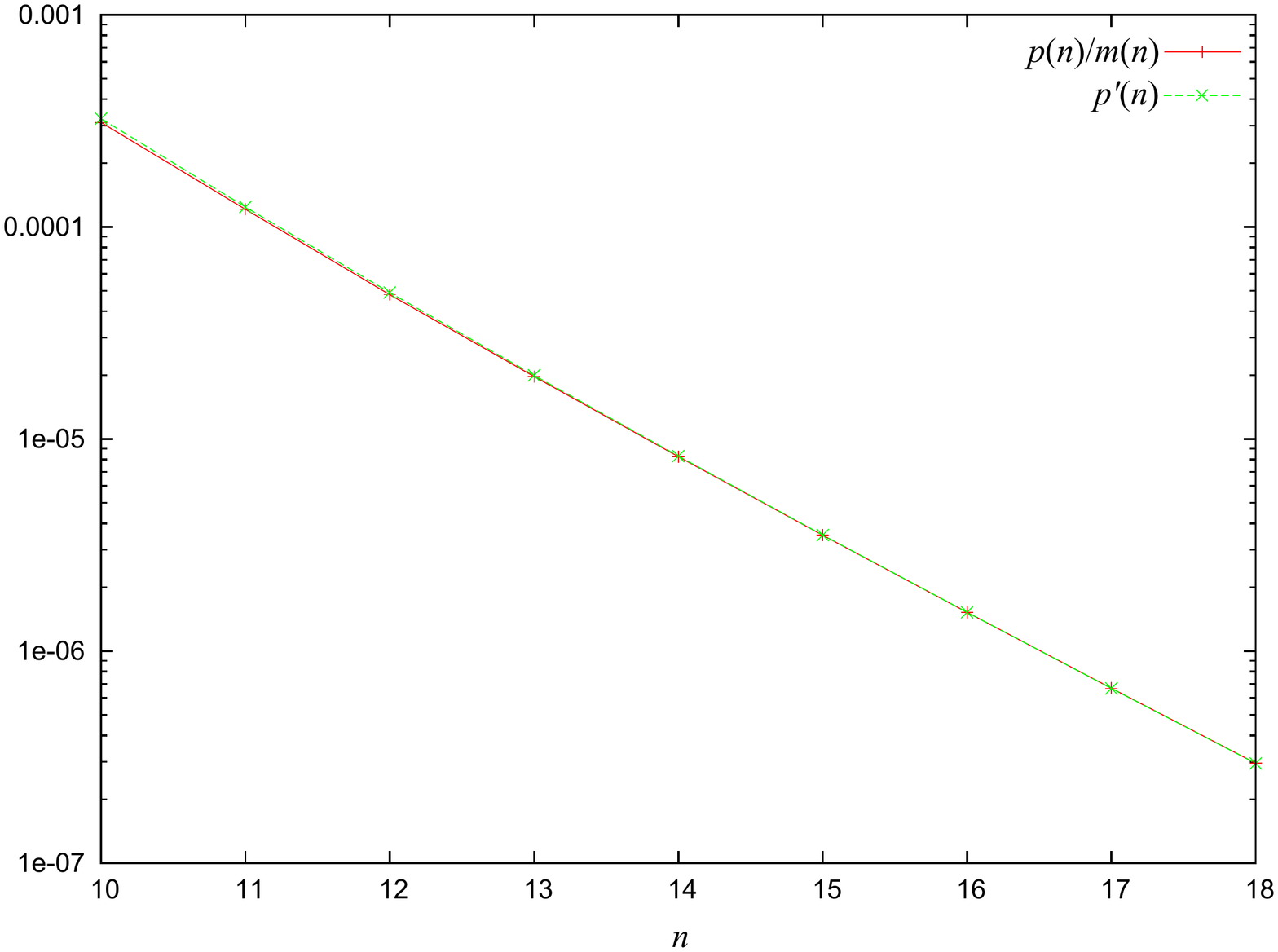}
\caption{Comparison of the predicted ($p(n)/m(n)$; solid line) and observed ($p'(n)$; dotted line) values of the required level of catalysis in the template-based version of the binary polymer CRS model, for various $n$,  using a log scale.}
\label{fig:comparelog}
\end{figure}

With the simulation results and the analytically calculated values of $m(n)$ from Table \ref{tab:all}, we can finally compare the predicted values $p(n)/m(n)$ with the observed values $p'(n)$, which is shown in Fig. \ref{fig:compare}. As this figure shows, there is a very close match between the predicted and observed values, with increasing accuracy for increasing values of $n$. Figure \ref{fig:comparelog} shows the same results on a log scale, to show  the results (and increasingly close fit) more clearly for larger values of $n$. This confirms our proposition that the required levels of catalysis in both model versions can be accurately predicted from each other using the molecule-reaction template match probabilities $m(n)$.

\begin{figure}[htb]
\centering
\includegraphics[width=8cm, angle=-90]{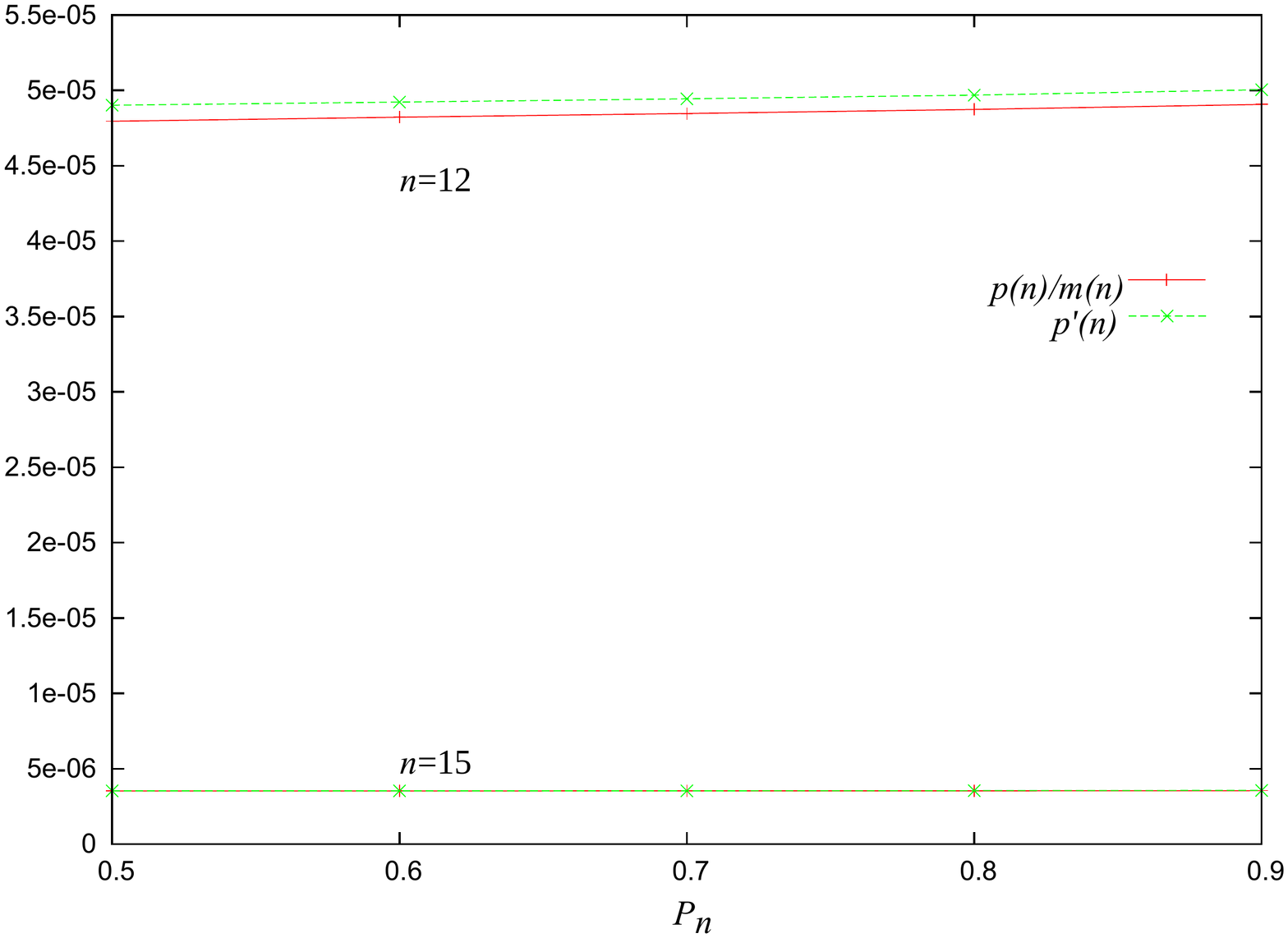}
\caption{Comparison of the predicted ($p(n)/m(n)$; solid line) and observed ($p'(n)$; dotted line) values of the required level of catalysis in the template-based version of the binary polymer CRS model, for various values of $P_n$.}
\label{fig:P_n}
\end{figure}

Next, we performed the same comparison for a fixed value of $n$ but varied the probability $P_n$ of finding RAF sets. For two given maximum molecule lengths ($n=12$ and $n=15$), we identified the values of $p(n)$ and $p'(n)$ to get probabilities of finding RAF sets of $P_n \in \{0.5,0.6,0.7,0.8,0.9\}$. Figure \ref{fig:P_n} shows the predicted values $p(n)/m(n)$ and the observed values $p'(n)$ for the template-based version of the model. Again, there is a close agreement (for $n=12$, the difference is about 2\%; for $n=15$, merely 0.3\%). So, the relationship also seems to hold for different values of $P_n$.

With these results, we can conclude to have established a direct relationship between two structurally distinct versions of the binary polymer model. More precisely, we have shown how these two models lead to identical predictions concerning the probability $P_n$ of finding autocatalytic sets, when the probabilities of catalysis are related by the $m(n)$ scaling factor, which can be calculated analytically.

\section{Conclusions and discussion} \label{sec:conclusion}

We have compared two different versions of a well-known binary polymer model of catalytic reaction systems as introduced in \cite{Kauffman:86,Kauffman:93} and investigated in detail in our own previous work \cite{Hordijk:11, Hordijk:04,Mossel:05,Steel:00}. In the first version of the model, the probability $p$ that an arbitrary molecule will catalyze an arbitrary reaction is equal and independent for each possible molecule-reaction pair. In the second version, a molecule will catalyze a reaction with similar probability $p'$ {\it only} if it matches the four-bit template around the reaction site.

At first glance, these two versions seem to impose rather different constraints, resulting in rather different (actual) molecule-reaction catalysis sets $C$. However, we have shown that it is possible to calculate analytically a factor $m$ which directly (and accurately) relates the catalysis probabilities $p$ and $p'$ in terms of the probabilities $P_n$ of finding autocatalytic sets in instances of both model versions. In other words, given the required level of catalysis in one model version to find autocatalytic sets with a given probability $P_n$, it is possible to accurately predict the required level of catalysis in the other model version to get the same probability $P_n$ of finding autocatalytic sets.

Because of this mathematical invariance between the two model versions, we can conclude that results obtained with the original (purely random and less realistic) model can still be considered relevant in the more realistic context of template-based catalysis. We have shown that the mathematical relationship holds for various $n$ (with fixed $P_n$) as well as for various $P_n$ (with fixed $n$), with increasing accuracy for increasing $n$, so it appears to be a very robust relationship. Furthermore, the analytical method for calculating the scaling factors $m(n)$ can be easily generalized to, for example, non-binary molecules or different template lengths, and  therefore does not depend on specific values of the model parameters.

This result clearly (and substantially) counters one potential and important objection that the binary polymer model in its original form (with purely randomly assigned catalysis) is too simplistic. We already showed earlier that it is straightforward to add more realistic extensions to the original model, such as template-based catalysis, which still results in similar behavior (in particular, a linear growth rate in the level of catalysis is sufficient for a high probability of finding RAF sets) \cite{Hordijk:11}. In the current paper, we  established a different, more direct and analytical relationship between the required levels of catalysis in these two model versions. Furthermore, it is interesting to note that recently, an {\em in-vitro} ``implementation'' of the binary polymer model was realized \cite{Taran:10}, where the food set consists of dinucleotides such as {\tt CA} and {\tt TG}, which can be interpreted as $0$s and $1$s in the binary model \cite{Kiedrowski:11}. The authors conclude that {\it ``different building blocks can be incorporated into the strand, promoting the formation of combinatorial libraries of oligonucleotides long enough to be folded into specific catalytically active structures and to potentially form initial autocatalytic sets''} \cite{Taran:10}. This is a further indication that the binary polymer model (and its results) has direct (bio)chemical relevance, including in the context of origin of life studies.

\section*{Acknowledgments}
The computations were performed at the Vital-IT ({\tt http://www.vital-it.ch}) Center for high-performance computing of the Swiss Institute of Bioinformatics.
MS thanks the Allan Wilson Centre for Molecular Ecology and Evolution and the Royal Society of NZ for funding.

\bibliographystyle{abbrv}
\bibliography{tmp-based}

\begin{thebibliography}{10}

\bibitem{wxMaxima}
{\tt http://wxmaxima.sourceforge.net}.

\bibitem{Benkoe:09}
G.~Benk{\"o}, F.~Centler, P.~Dittrich, C.~Flamm, B.~Stadler, and P.~F. Stadler.
\newblock A topological approach to chemical organizations.
\newblock {\em Alife}, 15:71--88, 2009.

\bibitem{Ganti:97}
T.~G\'{a}nti.
\newblock Biogenesis itself.
\newblock {\em Journal of Theoretical Biology}, 187:583--593, 1997.

\bibitem{Hordijk:10}
W.~Hordijk, J.~Hein, and M.~Steel.
\newblock Autocatalytic sets and the origin of life.
\newblock {\em Entropy}, 12(7):1733--1742, 2010.

\bibitem{Hordijk:11}
W.~Hordijk, S.~A. Kauffman, and M.~Steel.
\newblock Required levels of catalysis for emergence of autocatalytic sets in
  models of chemical reaction systems.
\newblock {\em International Journal of Molecular Sciences}, 12(5):3085--3101,
  2011.

\bibitem{Hordijk:04}
W.~Hordijk and M.~Steel.
\newblock Detecting autocatalytic, self-sustaining sets in chemical reaction
  systems.
\newblock {\em Journal of Theoretical Biology}, 227(4):451--461, 2004.

\bibitem{Jaramillo:10}
S.~Jaramillo, R.~Honorato-Zimmer, U.~Pereira, D.~Contreras, B.~Reynaert,
  V.~Hern{\'a}ndez, J.~Soto-Andrade, M.~C{\'a}rdenas, A.~Cornish-Bowden, and
  J.~Letelier.
\newblock {(M,R) Systems and RAF Sets: Common Ideas, Tools and Projections}.
\newblock In {\em Proceedings of the Alife XII Conference}, pages 94--100,
  2010.

\bibitem{Kauffman:86}
S.~A. Kauffman.
\newblock Autocatalytic sets of proteins.
\newblock {\em Journal of Theoretical Biology}, 119:1--24, 1986.

\bibitem{Kauffman:93}
S.~A. Kauffman.
\newblock {\em The Origins of Order}.
\newblock Oxford University Press, 1993.

\bibitem{Letelier:11}
J.-C. Letelier, M.~L. C{\'a}rdenas, and A.~Cornish-Bowden.
\newblock {From {\em L'Homme Machine} to metablic closure: steps towards
  understanding life}.
\newblock {\em Journal of Theoretical Biology}, 286:100--113, 2011.

\bibitem{Letelier:06}
J.-C. Letelier, J.~Soto-Andrade, F.~G. Abarz\'{u}a, A.~Cornish-Bowden, and
  M.~L. C\'{a}rdenas.
\newblock {Organizational invariance and metabolic closure: Analysis in terms
  of ({\em M}, {\em R}) systems}.
\newblock {\em Journal of Theoretical Biology}, 238:949--961, 2006.

\bibitem{Mossel:05}
E.~Mossel and M.~Steel.
\newblock Random biochemical networks: The probability of self-sustaining
  autocatalysis.
\newblock {\em Journal of Theoretical Biology}, 233(3):327--336, 2005.

\bibitem{Rosen:91}
R.~Rosen.
\newblock {\em Life Itself}.
\newblock Columbia University Press, 1991.

\bibitem{Sharov:91}
A.~Sharov.
\newblock Self-reproducing systems: structure, niche relations and evolution.
\newblock {\em BioSystems}, 25:237--249, 1991.

\bibitem{Stanley:86}
R.~P. Stanley.
\newblock {\em Enumerative Combinatorics}, volume~1.
\newblock Wadsworth \& Brooks/Cole, 1986.

\bibitem{Steel:00}
M.~Steel.
\newblock The emergence of a self-catalysing structure in abstract
  origin-of-life models.
\newblock {\em Applied Mathematics Letters}, 3:91--95, 2000.

\bibitem{Taran:10}
O.~Taran, O.~Thoennessen, K.~Achilles, and G.~von Kiedrowski.
\newblock Synthesis of information-carrying polymers of mixed sequences from
  double stranded short deoxynucleotides.
\newblock {\em Journal of Systems Chemistry}, 1(9), 2010.

\bibitem{Kiedrowski:11}
G.~von Kiedrowski, 2011.
\newblock Personal communication.

\end{thebibliography}

\end{document}